\documentclass[11pt]{article}
\usepackage{hyperref}
\pdfoutput=1
\begin{document}
\title{Leaf Roll-Up and Aquaplaning in Strong Winds and Floods}
\author{Laura Miller, Gregory Herschlag, and Arvind Santhanakrishnan\\
\\\vspace{6pt} Department of Mathematics, The University of North Carolina at Chapel Hill \\Chapel Hill, NC 27599, USA}
\maketitle
\begin{abstract}
\noindent The submitted fluid dynamics video (see \href{http://ecommons.library.cornell.edu/bitstream/1813/11489/1/miller_dfd_2008.mpg}{\underline{larger}} and \href{http://ecommons.library.cornell.edu/bitstream/1813/11489/2/miller_dfd_2008_small.mpg}{\underline{smaller}} sized versions) shows roll-up and aquaplaning in tree and vine leaves in both strong winds and water flows. The specific goal of this work is to determine how leaves roll up into drag reducing shapes during storms and aquaplane above the water surface during floods. In addition to gaining insight into mechanical adaptation in the natural world, this project might also inspire innovation in the engineering of structures and underwater vehicles.
\end{abstract}

\section{Summary}
Flexible plants, fungi, and sessile animals are thought to reconfigure in the wind and water to reduce the drag forces that act upon them. In strong winds, for example, leaves roll up into cone shapes that reduce flutter and drag when compared to paper cut-outs with similar shape and flexibility \cite{Vogel89}. Simple mathematical models of a flexible beam immersed in a two-dimensional flow will also exhibit this behavior \cite{Alben02}. During flash floods, herbaceous broad leaves aquaplane on the surface of the water which reduces drag. Vogel \cite{Vogel06} found that aquaplaning occurred in all ten herbaceous species and did not occur in all six tree species that he considered. What is less understood is how the mechanical properties of a two-dimensional leaf in a three-dimensional flow will passively allow roll up and aquaplaning. The larger goal of this work is to determine which morphological and mechanical features of organisms reduce the drag forces acting upon them in the water and during strong winds. 

\section{Methods}
Video imaging was used as the primary diagnostic technique to examine the leaf orientation under different conditions of fluid flow. The experimental portion of this work was conducted at the University of North Carolina at Chapel Hill, in the facilities of the Applied Math and Marine Science Fluids Lab and the Kier Lab in the Department of Biology. Hot wire anemometry was used to measure the average wind and water speeds.\\


\begin{thebibliography}{99}
\bibitem{Vogel89} S. Vogel, {\em Drag and Reconfiguration of Broad Leaves in High Winds}, J. Exp. Botany, 40, 941--948, 1989.
\bibitem{Alben02} S. Alben and M. Shelley and J. Zhang, {\em Drag Reduction through Self-Similar Bending of a Flexible Body}, Nature, 479--481, 2002.
\bibitem{Vogel06} S. Vogel, {\em Drag reduction by leaf aquaplaning in \textit{Hexastylis} (Aristolochiaceae) and other plant species in floods}, J. N. Am. Benthol. Soc., 25, 2--8, 2006.
\end{thebibliography}
\end{document}